\documentclass[runningheads]{llncs}

\usepackage[T1]{fontenc}
\PassOptionsToPackage{hyphens}{url}
\usepackage{url}
\usepackage[utf8]{inputenc}
\usepackage{graphicx}
\usepackage{amsmath}
\usepackage{booktabs}
\usepackage[hidelinks]{hyperref}

\graphicspath{{./figures/}}

\begin{document}

\title{Multi-Scale Spatio-Temporal Exploratory Analysis of Urban Vehicle Theft}
\titlerunning{Spatio-Temporal Analysis of Urban Vehicle Theft}

\author{Chaitanya Jobanputra\inst{1} \and
George Mohler\inst{2} \and
Sebastian Vollmer\inst{1} \and
Gerrit Gro{\ss}mann\inst{1}}

\authorrunning{C. Jobanputra et al.}

\institute{German Research Center for Artificial Intelligence (DFKI), Kaiserslautern, Germany\\
\email{\{Chaitanya\_rajeshbhai.Jobanputra, Sebastian.Vollmer, Gerrit.Grossmann\}@dfki.de}
\and
Boston College, Chestnut Hill, MA, USA\\
\email{mohlerg@bc.edu}}

\maketitle

\begingroup
\renewcommand\thefootnote{}
\footnotetext{Preprint. Under review.}
\endgroup

\begin{abstract}
Vehicle theft is a persistent urban problem with large economic and social costs. We analyze 521{,}098 stolen-vehicle reports filed with the Los Angeles Police Department from 2000 to 2022, at scales from hours to decades and from 100\,m grid cells to census tracts. Much of the work lies in making the administrative records fit for analysis: we separate thefts from recoveries and traffic reports filed in the same records, repair time fields, validate the geocoding, and harmonize census data across changing tract boundaries. Theft is strongly and persistently concentrated: 10.8\% of 250\,m cells hold half of all thefts, and six census tracts are significant rate hotspots in all 23 years after false discovery rate control. Thefts close in space and time are more frequent than expected by chance, but about half of this excess comes from events at identical coordinates. Lower-income tracts have more thefts at equal exposure, monthly temperature shows no association with theft, and thefts peak in the evening. From 2015--2019 to 2021--2022, the share of Kia and Hyundai among stolen vehicles rose from 1.5\% to 9.5\% and from 1.6\% to 7.7\%, mostly among model years 2011--2021, which largely lacked a standard immobilizer. All findings are descriptive. We also describe how the data quality issues we corrected would otherwise distort several of these patterns.

\end{abstract}

\section{Introduction}
\label{sec:intro}

More than one million vehicles were reported stolen in the United States in 2022, the first time since 2008, and California recorded the largest number of any state~\cite{nicb2023thefts}. Vehicle theft causes direct financial losses, but also feeds other offenses and absorbs police resources~\cite{clarke1992auto}. Like other crimes, it is spatially concentrated~\cite{sherman1989hot,weisburd2015law}, which makes place-based prevention attractive~\cite{braga2019hot}.

Many empirical studies of vehicle theft cover a few years and a single scale of analysis. Long administrative records allow different questions: whether hotspots persist over decades, how theft responds to changes in the vehicle fleet such as the recent Kia and Hyundai vulnerability~\cite{nhtsa2023kia,hldi2023hyundaikia}, and whether short-term near-repeat patterns and long-term concentration describe the same places. Such records also bring problems of their own. The files we analyze mix several kinds of records, change their formats over time, and lack coordinates for some years, and each of these issues can create apparent patterns.

We study vehicle theft in the City of Los Angeles, 2000--2022, using records of the Los Angeles Police Department (LAPD). Our contributions are:
\begin{enumerate}
\item A documented preparation of a 23-year administrative record: separating theft reports from other records, repairing time fields, validating geocoding, and harmonizing four census vintages on three tract geographies (Section~\ref{sec:data}).
\item A temporal analysis from hours to decades, with change points, seasonality and a test of the temperature hypothesis (Section~\ref{sec:temporal}).
\item A spatial and space-time analysis with explicit control of multiple testing: concentration, hotspots, socioeconomic correlates, near-repeat patterns and long-run persistence (Sections~\ref{sec:spatial} and~\ref{sec:spacetime}).
\item An analysis of vehicle targeting, including the Kia and Hyundai theft wave and its geography (Section~\ref{sec:vehicles}).
\end{enumerate}
The analysis is exploratory and descriptive; we make no causal claims. The code is available at \url{https://github.com/gerritgr/car_theft}; the crime records themselves are not public.

\section{Related Work}
\label{sec:related}

\paragraph{Crime concentration.}
Sherman et al.~\cite{sherman1989hot} showed that a small fraction of addresses accounts for a large share of police calls, and Weisburd~\cite{weisburd2015law} formulated this regularity as the law of crime concentration. Following street segments over 14 years, Weisburd et al.~\cite{weisburd2004trajectories} found that crime concentrations at micro places are generally stable over time. Hotspot mapping with kernel density estimation (KDE) and local statistics such as the Getis--Ord $G_i^*$~\cite{getis1992analysis,ord1995local} is a standard tool for describing it~\cite{chainey2008utility}. Andresen~\cite{andresen2006crime} showed that the choice of denominator changes which areas appear as high-crime areas, a point that matters for vehicle theft, where the population at risk is the vehicles present rather than the residents. Results also depend on the spatial units used, the modifiable areal unit problem (MAUP)~\cite{openshaw1984maup}.

\paragraph{Vehicle theft.}
Clarke and Harris~\cite{clarke1992auto} reviewed motives and prevention of auto theft, distinguishing joyriding, transportation and profit. Rice and Smith~\cite{rice2002socioecological} combined routine activity and social disorganization theories to explain auto theft rates across face blocks. Walsh and Taylor~\cite{walsh2007community} tested whether earlier community-level predictors of vehicle theft replicate in one Midwestern city (1990--2001) and found that only community socioeconomic status and theft rates in nearby communities were consistently significant. Mao et al.~\cite{mao2018spacetime} found a temporally stable pattern of vehicle theft, including thefts of non-motor vehicles, in a district of Shanghai. For near repeats, a pattern first established for burglary~\cite{townsley2003infectious,johnson2004burglary,johnson2007space}, Lockwood~\cite{lockwood2012presence} and Block and Fujita~\cite{block2013patterns} reported near-repeat patterns of motor vehicle theft. Self-exciting point processes model such short-term contagion directly~\cite{mohler2011self}.

\paragraph{Temporal patterns.}
Property crime often peaks in summer, but seasonality varies across space~\cite{andresen2013seasonality}, and cities with stronger seasonal weather variation show more distinct summer increases~\cite{linning2017crime}. Routine activity theory~\cite{cohen1979social} explains such patterns by changes in the convergence of offenders, targets and guardians.

\paragraph{Kia and Hyundai.}
Passive immobilizers were standard on only 26\% of 2015 Kia and Hyundai vehicle series, against 96\% for all other manufacturers combined~\cite{hldi2023hyundaikia}. Theft claim frequency for 2003--2023 Kia and Hyundai models was more than ten times higher in the first half of 2023 than in the first half of 2020~\cite{hldi2023hyundaikia}. After theft methods for these vehicles spread on social media, the manufacturers launched a software campaign in February 2023~\cite{nhtsa2023kia}. Using LAPD data, Brantingham et al.~\cite{brantingham2026forecasting} analyzed thefts of the vulnerable Kia and Hyundai models in Los Angeles and forecast elevated thefts for at least two decades.

\paragraph{Positioning.}
We combine these strands in one long record, analyzed at several spatial and temporal scales with consistent definitions. Compared with most hotspot studies, we put more weight on data preparation and on how preparation choices affect spatial results, and we control multiple testing throughout.

\section{Data and Preprocessing}
\label{sec:data}

\subsection{Records and Event Definition}

The data are LAPD vehicle-related crime records, one file per year from 2000 to 2022, with 5{,}000{,}044 rows. Each row describes a vehicle in a report, identified by the report number, and carries a crime class, the occurrence date and time, an address, a reporting district and, in most years, stored coordinates. The files contain six groups of crime classes (Table~\ref{tab:dataset}). Only class 510 records stolen vehicles. Class 521 records recovered vehicles, with the date and place of the recovery, and class 997 records vehicles involved in traffic reports (flagged as such in the data). Counting all three as thefts would multiply the volume about sevenfold and mix theft locations with recovery and collision locations.

We therefore define an event as a class 510 report with an occurrence date in 2000--2022, counted once per report number; duplicate rows of the same report are merged, preferring the copy with stored coordinates. This yields 521{,}098 thefts, between 15{,}000 and 33{,}000 per year. A theft counts as recovered when a class 521 record carries the same report number; this holds for 87.4\% of thefts, with a median of 4--5 days to recovery. Most class 521 records have no matching theft report in the data, so recoveries cannot stand in for thefts.

\begin{table}[t]
\centering
\caption{Records by crime class (unique report and class, 2000--2022) and coordinate sources of the analyzed theft reports.}
\label{tab:dataset}
\small
\begin{tabular}{lrr}
\toprule
Category & Count & Share \\
\midrule
Vehicle theft (510) & 521{,}098 & 11.6\% \\
Vehicle recovered (521) & 2{,}447{,}625 & 54.5\% \\
Vehicle in traffic report (997) & 770{,}333 & 17.2\% \\
Burglary from vehicle (330) & 364{,}776 & 8.1\% \\
Vandalism (740, 745) & 228{,}963 & 5.1\% \\
Theft from vehicle (420, 331) & 158{,}132 & 3.5\% \\
\midrule
Thefts with stored LAPD coordinates & 377{,}984 & 72.5\% \\
Thefts geocoded with the Census geocoder & 127{,}808 & 24.5\% \\
Thefts geocoded with Nominatim & 4{,}226 & 0.8\% \\
Geocodes outside the city (rejected) & 2{,}782 & 0.5\% \\
Thefts without coordinates & 8{,}298 & 1.6\% \\
\bottomrule
\end{tabular}
\end{table}

\paragraph{Occurrence time.}
The time field changes format over the years: zero-padded until 2017, not zero-padded in 2018--2021, and absent in 2022, where every record carries the same placeholder. A time of 00:01 is a placeholder for an unknown time. We treat both as missing, so hour-of-day analyses use 491{,}028 thefts (94.2\%) and exclude 2022.

\paragraph{Incomplete end of record.}
The 2022 file ends on December 30, 2022, and its last weeks are incomplete: weekly counts fall from about 470 in October to about 300 in December, and December 2022 is the lowest month of all 23 years. Monthly analyses therefore end in October 2022; annual counts for 2022 are understated by roughly 6\%.

\subsection{Geocoding and Location Sources}
\label{subsec:geocoding}

\begin{figure}[t]
\centering
\includegraphics[width=0.95\linewidth]{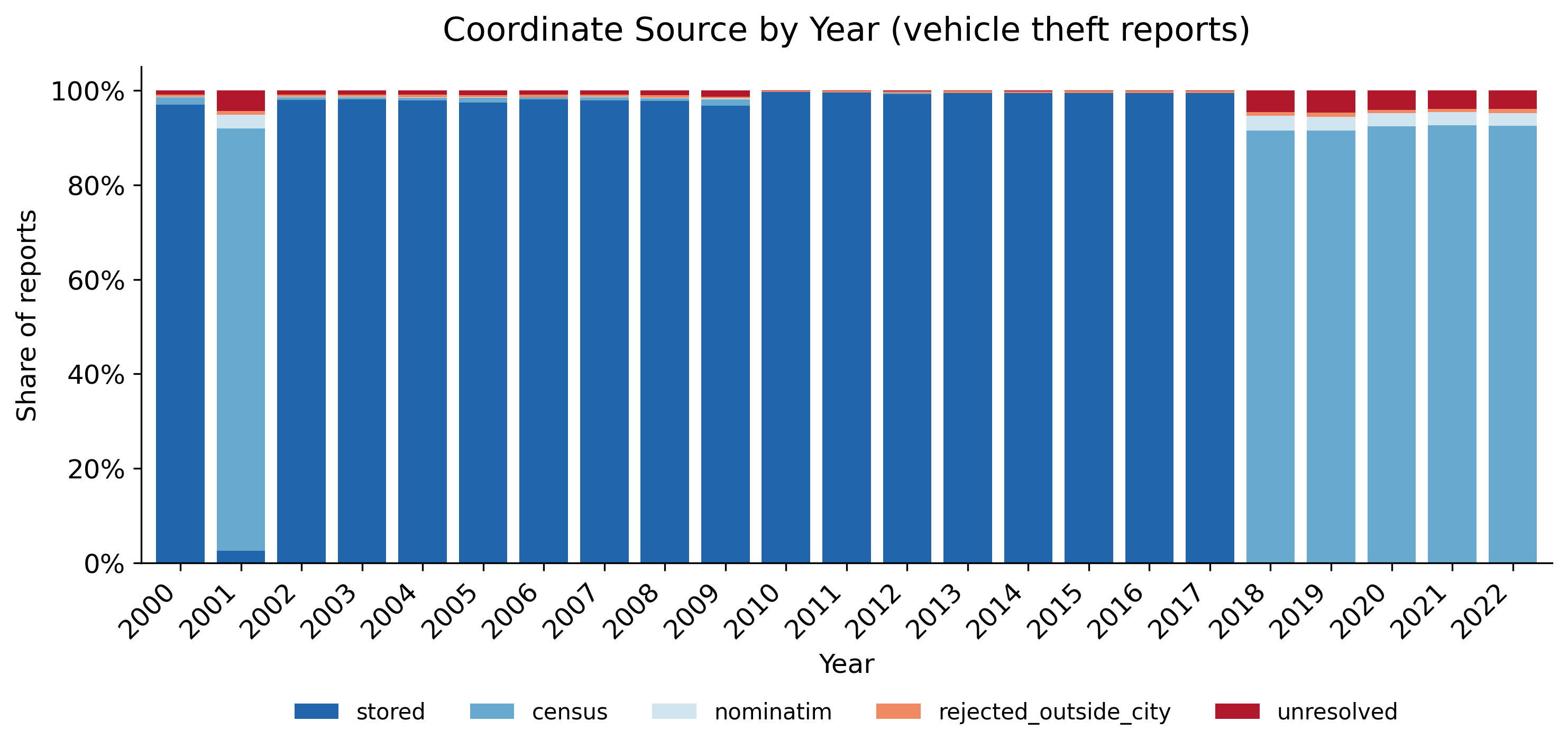}
\caption{Coordinate source of theft reports by year. In 2001 and 2018--2022, locations come almost entirely from address geocoding.}
\label{fig:coverage}
\end{figure}

Stored coordinates are available for most years but are almost absent in 2001 and 2018--2022 (Fig.~\ref{fig:coverage}). For reports without coordinates we geocoded the address with the Census Bureau batch geocoder~\cite{censusgeocoder} and, for addresses it could not match, with Nominatim~\cite{nominatim}, a geocoder for OpenStreetMap data, restricted to a box around the reporting district (or around the ZIP code area or the whole region when no reliable district center was available). Geocodes outside the City of Los Angeles cannot be LAPD incidents and are discarded. In total, 97.9\% of thefts have coordinates inside the city, well above the minimum acceptable hit rate of 85\% proposed by Ratcliffe~\cite{ratcliffe2004geocoding}. Because the location source changes abruptly between years, we treat 2001 and 2018--2022 as \emph{geocoded years} and check every spatial result that changes at these years against this switch.

Some reports appear twice in the raw files, once with stored coordinates and once without; the second copy was geocoded independently. For copies with the same street address, the distance between the two points measures the geocoding error (Table~\ref{tab:geocoding}), the positional component of geocoding quality, whose errors are often neither small nor random~\cite{zandbergen2009geocoding}. Census matches of numbered addresses lie a median of 12\,m from the stored point, and 94\% lie within 100\,m; Nominatim matches are much less precise. Two caveats limit this validation. Almost all validation pairs come from years with stored coordinates, so the accuracy of the geocoded years is inferred rather than measured. In addition, intersection addresses, 18.1\% of geocoded points in the geocoded years, could not be validated at all. Separately, we reverse-geocoded a stratified random sample of 540 stored coordinates (30 per year with stored coordinates): in 77.6\% the street name of the address matched the street at the coordinate.

\begin{table}[t]
\centering
\caption{Distance between geocoded and stored positions for copies of the same report with the same address (all crime classes), in meters.}
\label{tab:geocoding}
\small
\begin{tabular}{lrrrr}
\toprule
Geocoder and match & Pairs & Median & 90th percentile & Within 100\,m \\
\midrule
Census, exact match & 20{,}477 & 11.5 & 63.9 & 94.4\% \\
Census, non-exact match & 6{,}430 & 11.7 & 67.5 & 93.6\% \\
Nominatim & 354 & 94.8 & 2{,}995.8 & 50.6\% \\
\bottomrule
\end{tabular}
\end{table}

Reports without usable coordinates are not spread evenly: their share ranges from 0.8\% to 4.1\% across the 21 LAPD geographic areas, and reaches 12.4\% in the Harbor area during the geocoded years. It is not related to an area's theft volume (Spearman $\rho=0.25$, $p=0.27$).

\subsection{Study Area and Denominators}
\label{subsec:denominators}

LAPD covers the City of Los Angeles only. We use the 1{,}110 census tracts of 2020 with at least half of their area inside the city~\cite{tiger2022}, which contain 505{,}666 thefts. Tract-level denominators come from the Census 2000 Summary File 3~\cite{census2000sf3} and the American Community Survey (ACS) 5-year estimates for 2005--2009, 2012--2016 and 2018--2022~\cite{census2022acs}: aggregate vehicles available to households, median household income and population. Where the aggregate vehicle count is suppressed, we estimate it from the distribution of households by number of vehicles. The four vintages use three different tract geographies; we reallocate them to 2020 tracts with the Census relationship files~\cite{census2020relationship}, by population share from 2000 to 2010 and by land area share from 2010 to 2020. Annual values are interpolated linearly between the reference years of the vintages (2000, 2007, 2014 and 2020), and income is converted to 2022 dollars with the consumer price index for all urban consumers~\cite{bls_cpi}. The 14 special land-use tracts (airport, port and parks) are excluded from all rate-based analyses.

Household vehicles measure the vehicles of residents, not all vehicles parked in a tract. Rates per household vehicle are therefore high in commercial and transit areas with few residents, and we use them with this caveat.

\section{Methods}
\label{sec:methods}

\paragraph{Local and global spatial statistics.}
The Getis--Ord $G_i^*$ statistic~\cite{getis1992analysis,ord1995local} compares the sum of a variable over a tract and its neighbors with the citywide mean; large positive values indicate clusters of high values (hotspots, HH) and large negative values clusters of low values (coldspots, LL). We use contiguity weights including the tract itself, row-standardized, and conditional permutation tests with 9{,}999 permutations, converted to two-sided pseudo $p$-values (smallest attainable value $0.0002$). Because many tracts are tested at once, we control the false discovery rate (FDR) with the Benjamini--Hochberg (BH) procedure~\cite{benjamini1995controlling} and call a tract a hotspot if it is HH with an FDR-adjusted value $q\le0.05$. Global Moran's $I$~\cite{moran1950notes} measures overall spatial autocorrelation. Computations use PySAL~\cite{rey2007pysal}.

\paragraph{Kernel density.}
Density surfaces use a Gaussian kernel with a bandwidth of 500\,m~\cite{silverman1986density} on a fixed grid of 100\,m cells over the city. Each surface is normalized to sum to one, so surfaces of periods with different volumes are directly comparable; the share of density that moves between two periods is half the $L_1$ distance between their surfaces.

\paragraph{Space-time interaction.}
The Knox test~\cite{knox1964detection} counts pairs of events that are close in both space and time and compares the count with a permutation distribution. We use exact distances and draw a random sample of 5{,}000 thefts per year. Permuting dates across all years makes the test sensitive to changes in where events occur over the years~\cite{kulldorff1999knox}, including the switches of location source; we therefore permute dates only within each year.

\paragraph{Temporal decomposition and trends.}
The monthly series of thefts per day is decomposed with STL, a seasonal-trend decomposition based on locally weighted regression~\cite{cleveland1990stl}. Change points are detected on the deseasonalized series with the pruned exact linear time (PELT) algorithm~\cite{killick2012optimal,truong2020selective}, using a continuous piecewise linear cost, so that change points mark changes in slope, and a penalty based on the Bayesian information criterion. Trends in short annual series are tested with the Mann--Kendall test~\cite{mann1945nonparametric} with the Hamed--Rao correction for serial correlation~\cite{hamed1998modified}.

\paragraph{Regression.}
Tract theft counts are modeled with negative binomial (NB) regression~\cite{osgood2000poisson}, with exposure entered as covariates. We report standard errors clustered by LAPD area in addition to model-based ones. Tracts are spatially dependent, so all tract-level $p$-values should be read as optimistic.

\section{Temporal Patterns}
\label{sec:temporal}

\begin{figure}[t]
\centering
\includegraphics[width=0.95\linewidth]{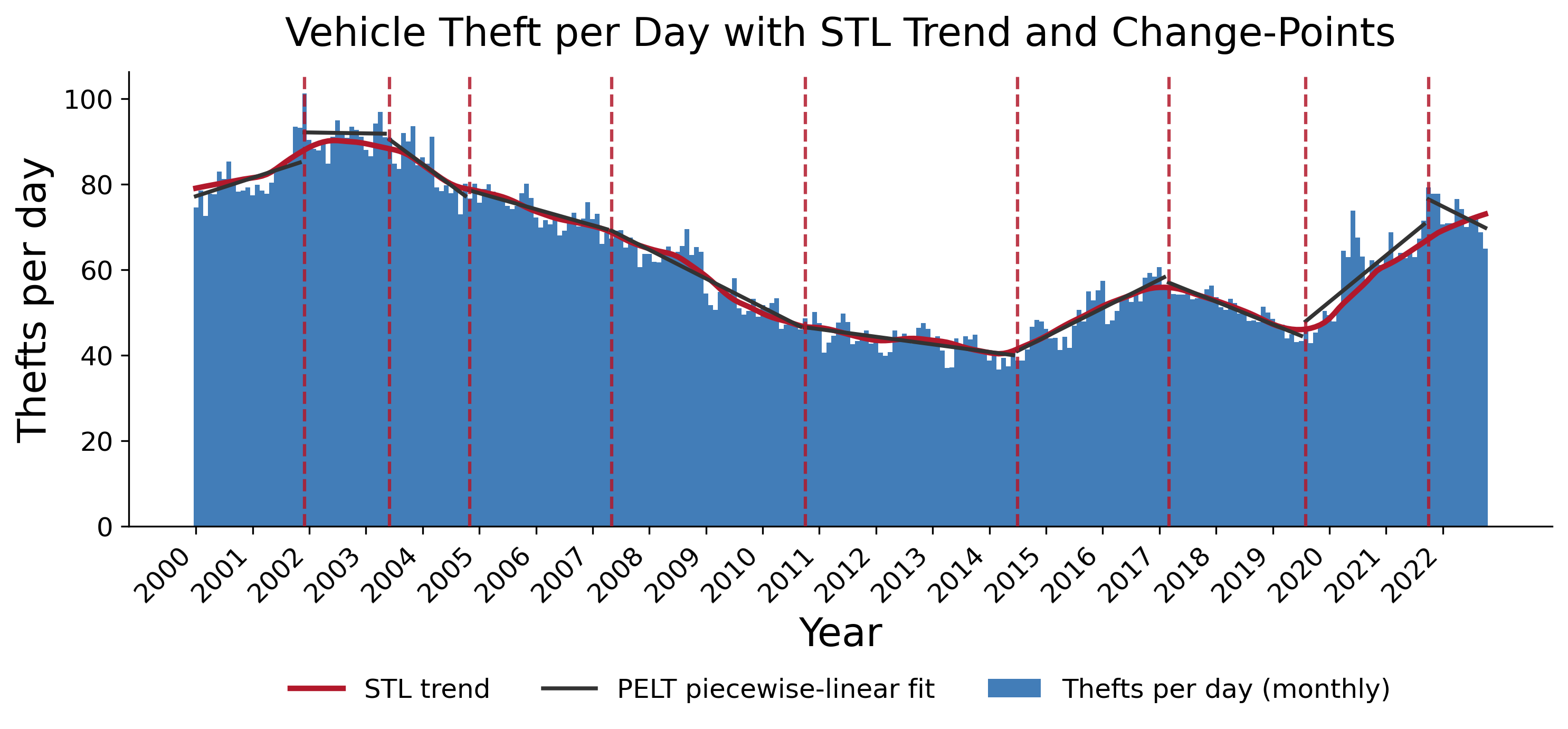}
\caption{Monthly thefts per day, STL trend, and the piecewise linear fit with change points detected by PELT (January 2000 to October 2022).}
\label{fig:trend}
\end{figure}

\paragraph{Long-run trend.}
Thefts declined from about 90 per day in 2002--2003 to about 41 per day in 2014, and then rose again in two waves (Fig.~\ref{fig:trend}). PELT detects nine changes in slope (Table~\ref{tab:changepoints}); depending on the penalty, between five and ten are found, so we interpret them as a description of the trend rather than as distinct events. The steepest increase began in August 2019, before the COVID-19 pandemic, and continued until late 2021.

\begin{table}[t]
\centering
\caption{Change points in the slope of the deseasonalized daily theft rate. Slopes are in thefts per day per year.}
\label{tab:changepoints}
\small
\begin{tabular}{lrrr}
\toprule
Change point & Level (thefts/day) & Slope before & Slope after \\
\midrule
Dec 2001 & 92.1 & $+4.29$ & $-0.22$ \\
Jun 2003 & 90.4 & $-0.22$ & $-9.87$ \\
Nov 2004 & 78.5 & $-9.87$ & $-3.75$ \\
May 2007 & 69.1 & $-3.75$ & $-6.73$ \\
Oct 2010 & 46.5 & $-6.73$ & $-1.78$ \\
Jul 2014 & 41.0 & $-1.78$ & $+6.66$ \\
Mar 2017 & 57.0 & $+6.66$ & $-5.33$ \\
Aug 2019 & 48.0 & $-5.33$ & $+10.87$ \\
Oct 2021 & 76.4 & $+10.87$ & $-6.64$ \\
\bottomrule
\end{tabular}
\end{table}

\paragraph{Seasonality and temperature.}
The seasonal component is small relative to the trend: averaged over the years, it ranges from $-2.1$ thefts per day in March to $+3.1$ in November, against a mean of about 62 thefts per day. This is consistent with the finding that cities with milder climates show weaker seasonality~\cite{linning2017crime}; thefts are highest in late autumn. Mean monthly temperature in downtown Los Angeles~\cite{openmeteo} is not associated with theft, neither across calendar months (Spearman $\rho=-0.07$, $p=0.83$, $n=12$) nor between the deviations from the trend and seasonal cycle and the temperature anomalies ($\rho=-0.03$, $p=0.67$, $n=274$).

\begin{figure}[t]
\centering
\includegraphics[width=0.9\linewidth]{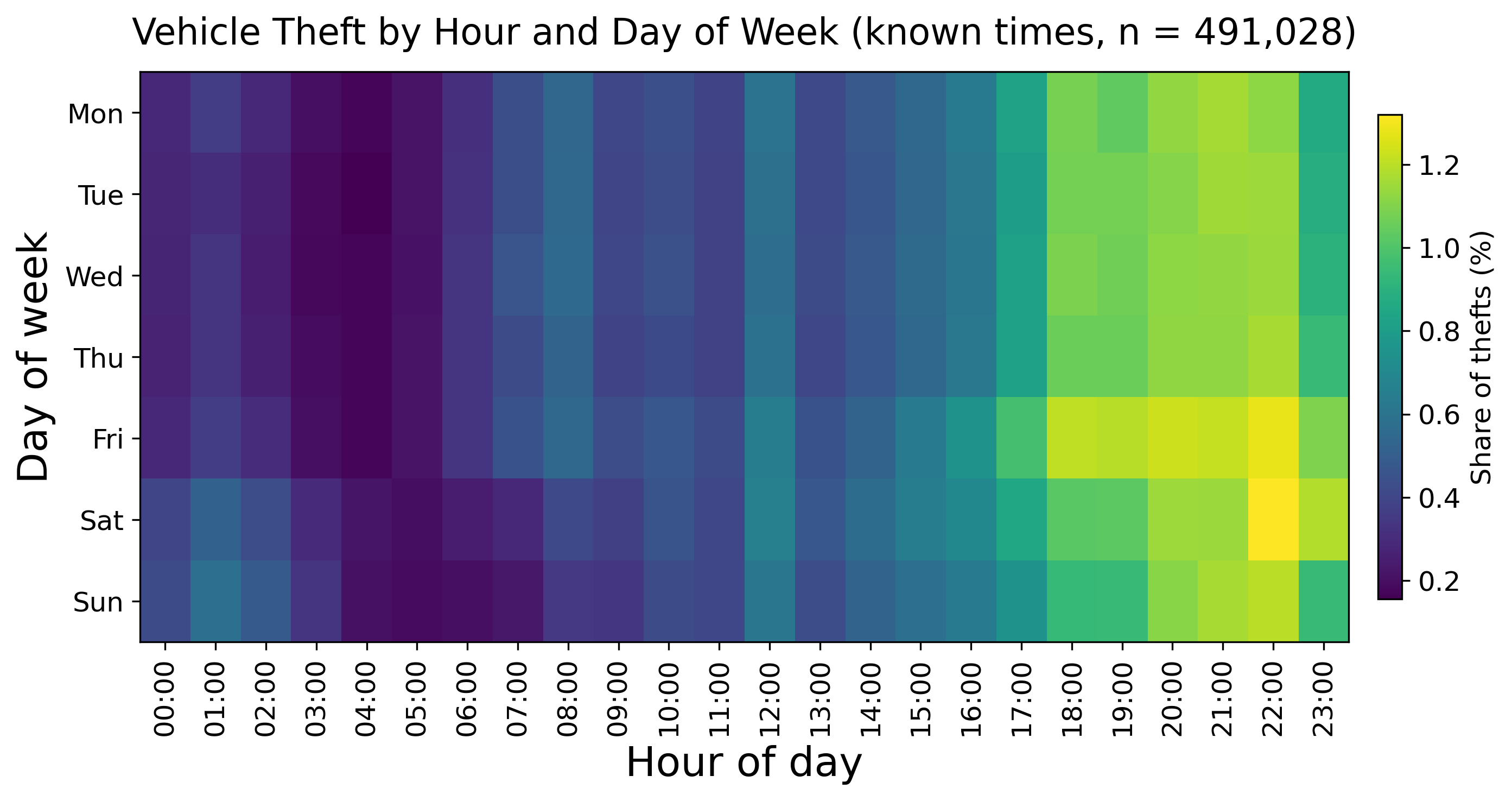}
\caption{Share of thefts by hour of day and day of week, 2000--2021, thefts with a known occurrence time.}
\label{fig:heatmap}
\end{figure}

\paragraph{Time of day.}
Thefts concentrate in the evening: 46.2\% of thefts with a known time occur between 18:00 and midnight, with a peak at 22:00, and only 7.6\% between 04:00 and 08:00 (Fig.~\ref{fig:heatmap}). Friday and Saturday evenings are the busiest. The excess at 12:00 (4.3\%, against about 3\% in the neighboring hours) likely reflects rounded times.

\section{Spatial Structure}
\label{sec:spatial}

\begin{figure}[t]
\centering
\includegraphics[width=0.75\linewidth]{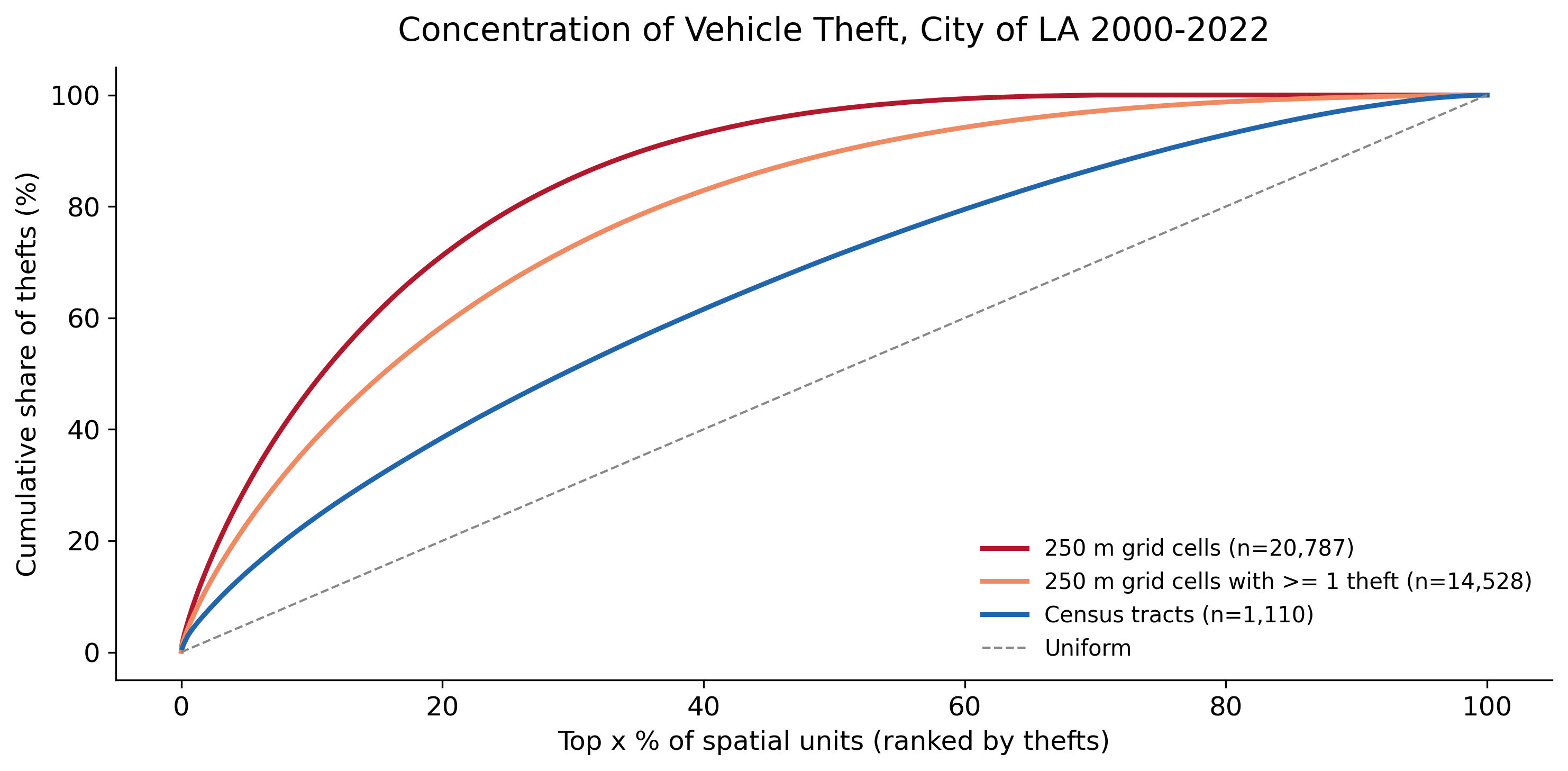}
\caption{Cumulative share of thefts in the top $x$\% of spatial units, for 250\,m grid cells, grid cells with at least one theft, and census tracts, 2000--2022.}
\label{fig:concentration}
\end{figure}

\paragraph{Concentration.}
Theft is strongly concentrated, and the degree of concentration depends on the spatial unit (Fig.~\ref{fig:concentration}). Half of all thefts fall in 10.8\% of the 250\,m grid cells of the city (15.5\% of the cells with at least one theft), but in 29.3\% of the census tracts. Tract-level results, which are needed to relate theft to census data, therefore understate concentration, an instance of the MAUP~\cite{openshaw1984maup}.

\begin{figure}[t]
\centering
\includegraphics[width=0.48\linewidth]{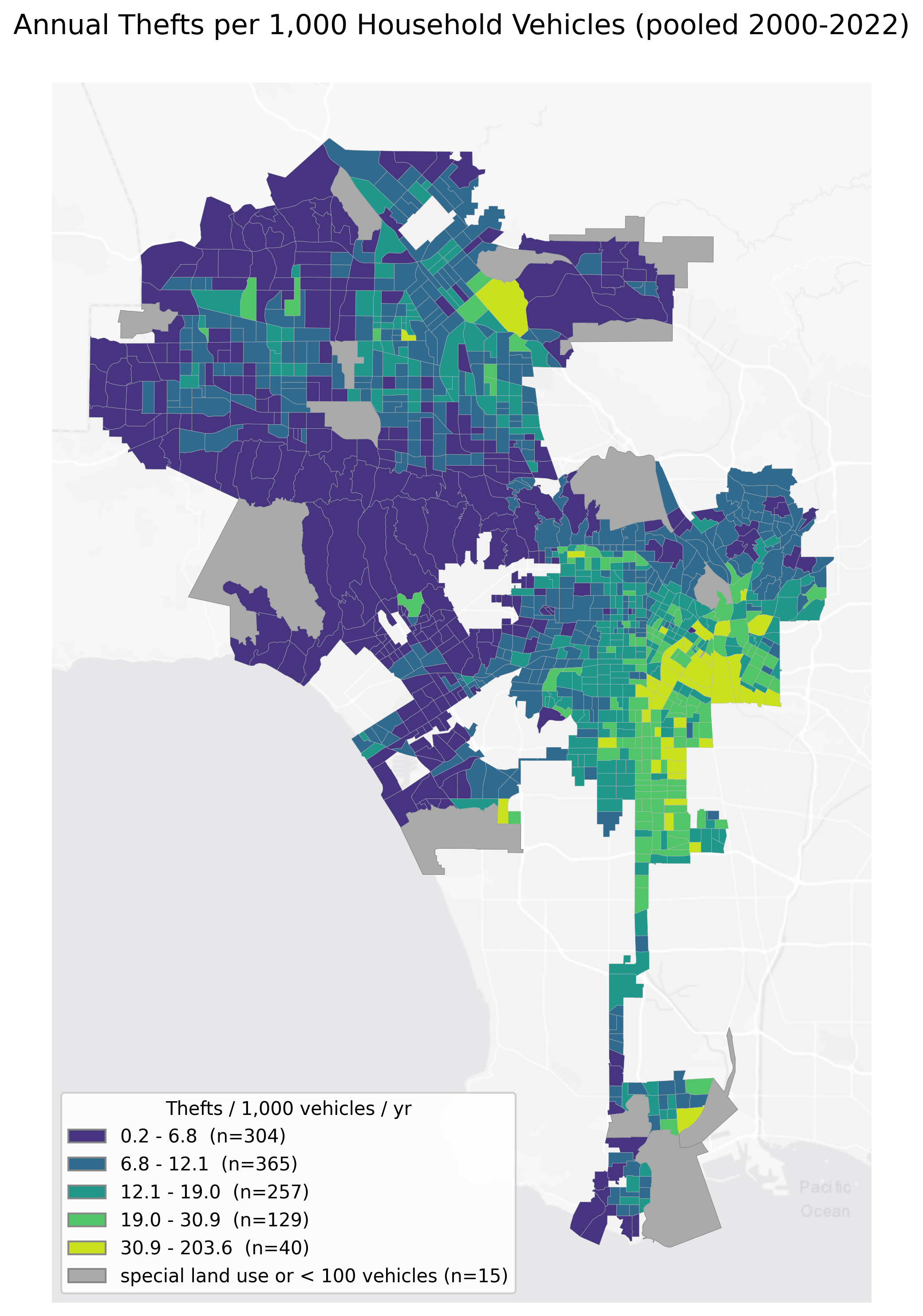}\hfill
\includegraphics[width=0.43\linewidth]{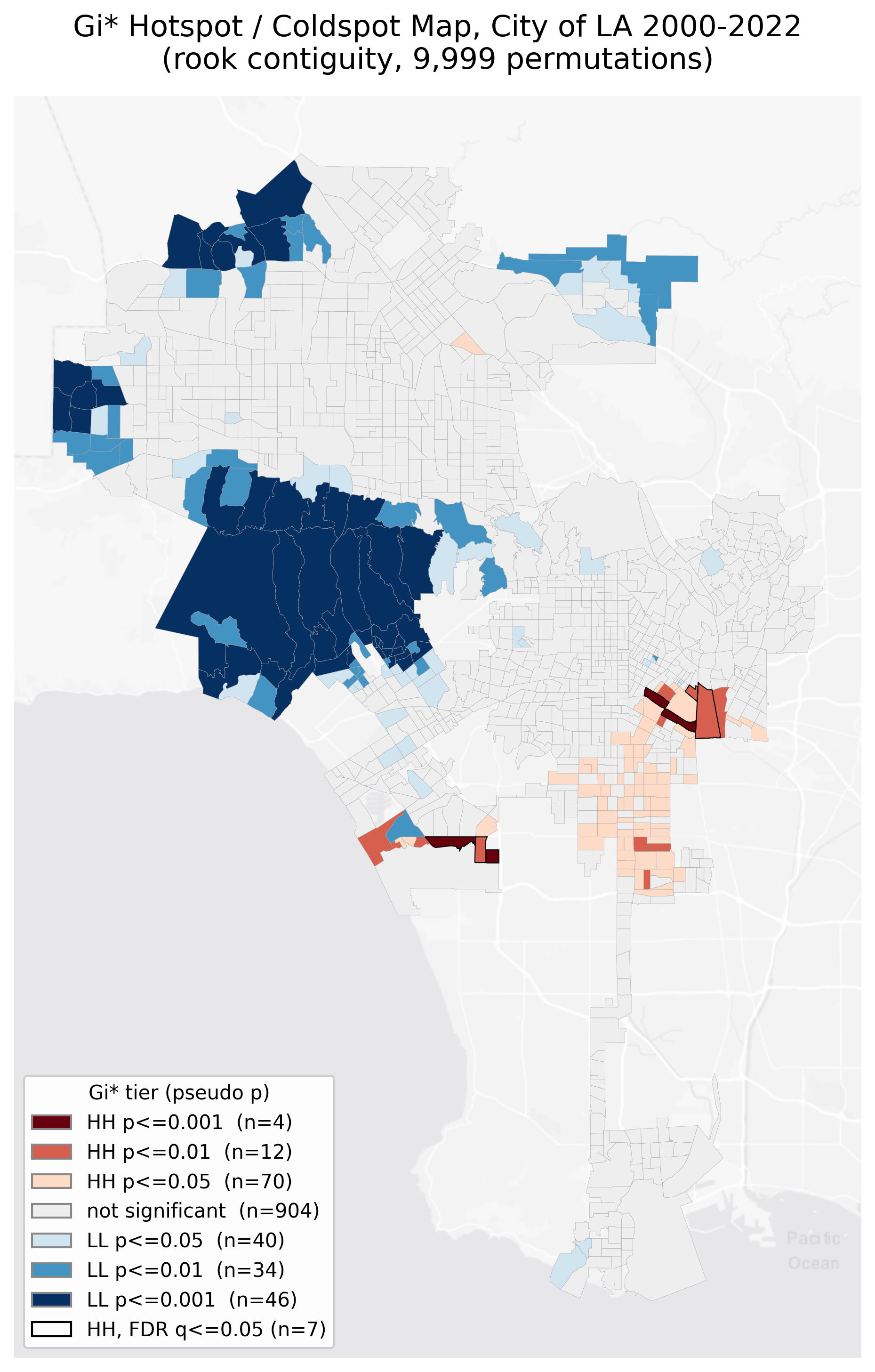}
\caption{Left: pooled annual thefts per 1{,}000 household vehicles by census tract, 2000--2022 (gray: special land use or fewer than 100 vehicles). Right: $G_i^*$ tiers of the 23-year theft count (rook contiguity, 9{,}999 permutations, two-sided pseudo $p$-values); outlined tracts are hotspots after FDR control.}
\label{fig:maps}
\end{figure}

\paragraph{Rates and hotspots.}
The pooled annual rate per 1{,}000 household vehicles has a median of 10.2 across tracts (interquartile range 6.3--15.4) and is highest south and southwest of Downtown, extending into South Los Angeles, and in the port area (Fig.~\ref{fig:maps}, left). For the cross-sectional hotspots we apply $G_i^*$ to the 23-year theft count, so they describe where thefts happen (volume), whereas the persistence analysis in Section~\ref{sec:spacetime} uses rates (risk per vehicle). With raw $p$-values, 86 tracts are HH at $p\le0.05$; after FDR control, seven tracts remain hotspots and 62 remain coldspots (Fig.~\ref{fig:maps}, right). The HH classification changes little with the choice of weights and variable: queen or rook contiguity combined with counts or rates classify 92.5--97.7\% of tracts in the same way as the main specification. Using only the years with stored coordinates, or omitting the Nominatim geocodes, yields exactly the same seven hotspots.

The strongest hotspot, tract 2774.00 next to Los Angeles International Airport (LAX), has 22.4\% of its thefts at a single coordinate, against a median of 5.3\% across tracts, and nearby tracts have unusually new stolen vehicles of makes common in rental fleets. This hotspot probably reflects airport and rental parking rather than residential theft.

\paragraph{Redistribution over time.}
The geography of theft changed little. Between 2002--2009 and 2010--2017, both with stored coordinates, 7.6\% of the kernel density moved; between the two halves of 2010--2017 alone, 6.6\% moved. The comparison of 2010--2017 with the geocoded years 2018--2022 gives 8.5\%, which includes the effect of the change of location source.

\paragraph{Income and other correlates.}
Tracts with lower median income have more thefts. The rank correlation between income and the rate per household vehicle ($\rho=-0.79$) is much stronger than with the rate per resident ($\rho=-0.49$), because households in higher-income tracts own more vehicles ($\rho=0.89$ between income and vehicles per resident); the vehicle-based rate thus builds part of the income gradient into its denominator. An NB model of the 23-year theft count with log household vehicles, log population and log land area as covariates avoids this. Its income coefficient is $-1.17$ (standard error 0.05; clustered 0.16; Table~\ref{tab:nb}): at equal vehicles, population and area, a 10\% higher median income goes with about 10.6\% fewer thefts. Adding densities of points of interest (POIs) from OpenStreetMap~\cite{osm} barely changes the income coefficient ($-1.11$); transit stop density has a positive coefficient ($0.15$, clustered standard error 0.04). The POI data are a 2026 snapshot and serve as a description only. The model residuals remain spatially autocorrelated (Moran's $I=0.38$).

\begin{table}[t]
\centering
\caption{Negative binomial model of the 23-year theft count per tract ($n=1{,}095$ tracts). Standard errors (SE) are model-based and clustered by the 21 LAPD areas.}
\label{tab:nb}
\small
\begin{tabular}{lrrr}
\toprule
Covariate & Coefficient & SE & Clustered SE \\
\midrule
log household vehicles & 0.410 & 0.057 & 0.146 \\
log population & 0.066 & 0.063 & 0.112 \\
log land area & 0.306 & 0.023 & 0.064 \\
log median income & $-1.170$ & 0.050 & 0.158 \\
\bottomrule
\end{tabular}
\end{table}

\section{Space-Time Dynamics}
\label{sec:spacetime}

\paragraph{Clustering over time.}
Global Moran's $I$ of annual tract counts is significant in every year ($p\le0.001$) and rose from 0.20 in 2000 to 0.46 in 2022 (Mann--Kendall $\tau=0.70$, $p=0.004$). The increase also holds when only years with stored coordinates are used ($\tau=0.74$, $p<0.001$), so it is not an artifact of geocoding. Even as total volume fell and rose again, theft became more spatially clustered.

\paragraph{Near repeats.}
All twelve combinations of distance and time bands show more close pairs than expected (BH-adjusted $q\le0.001$; Table~\ref{tab:knox}), with the largest excess for pairs within 250\,m and 7 days (ratio 1.18). The excess decays with distance and time and is small beyond 1\,km. However, 30.1\% of the sampled thefts share their exact coordinates with another sampled theft. Excluding pairs at zero distance halves the excess at 250\,m and 7 days (ratio 1.08): about half of the short-range effect consists of events at identical coordinates, which can be repeats at the same address or several addresses that the geocoder or the stored coordinates place at one point.

\begin{table}[t]
\centering
\caption{Knox ratios (observed / mean permuted number of close pairs; 5{,}000 thefts per year, 999 within-year permutations). In parentheses: pairs at zero distance excluded.}
\label{tab:knox}
\small
\begin{tabular}{lrrr}
\toprule
Distance & 7 days & 30 days & 90 days \\
\midrule
250\,m & 1.178 (1.083) & 1.060 (1.030) & 1.028 (1.016) \\
500\,m & 1.087 (1.057) & 1.038 (1.029) & 1.017 (1.013) \\
1{,}000\,m & 1.044 & 1.023 & 1.010 \\
2{,}000\,m & 1.025 & 1.013 & 1.006 \\
\bottomrule
\end{tabular}
\end{table}

\begin{figure}[t]
\centering
\includegraphics[width=0.48\linewidth]{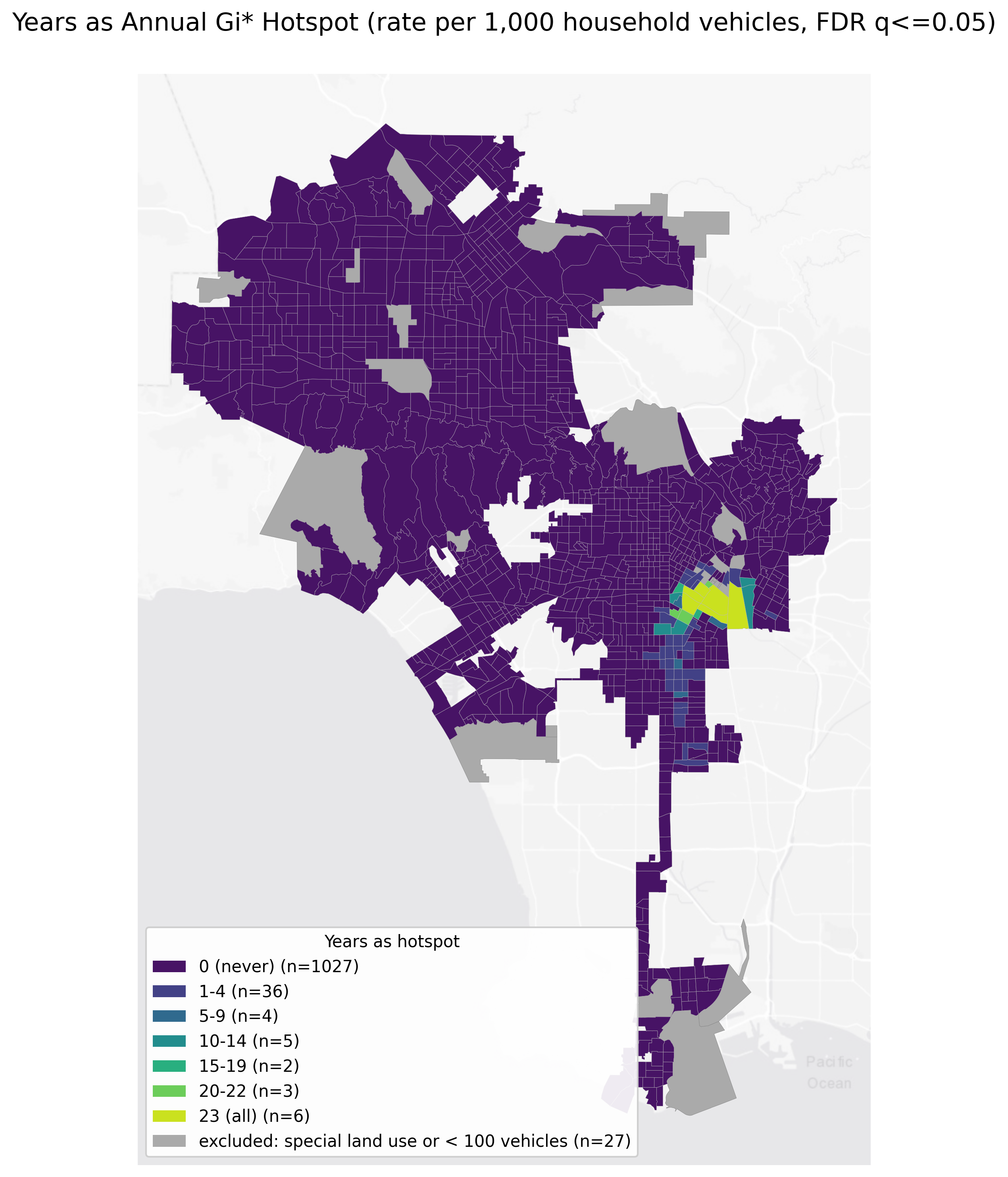}\hfill
\includegraphics[width=0.48\linewidth]{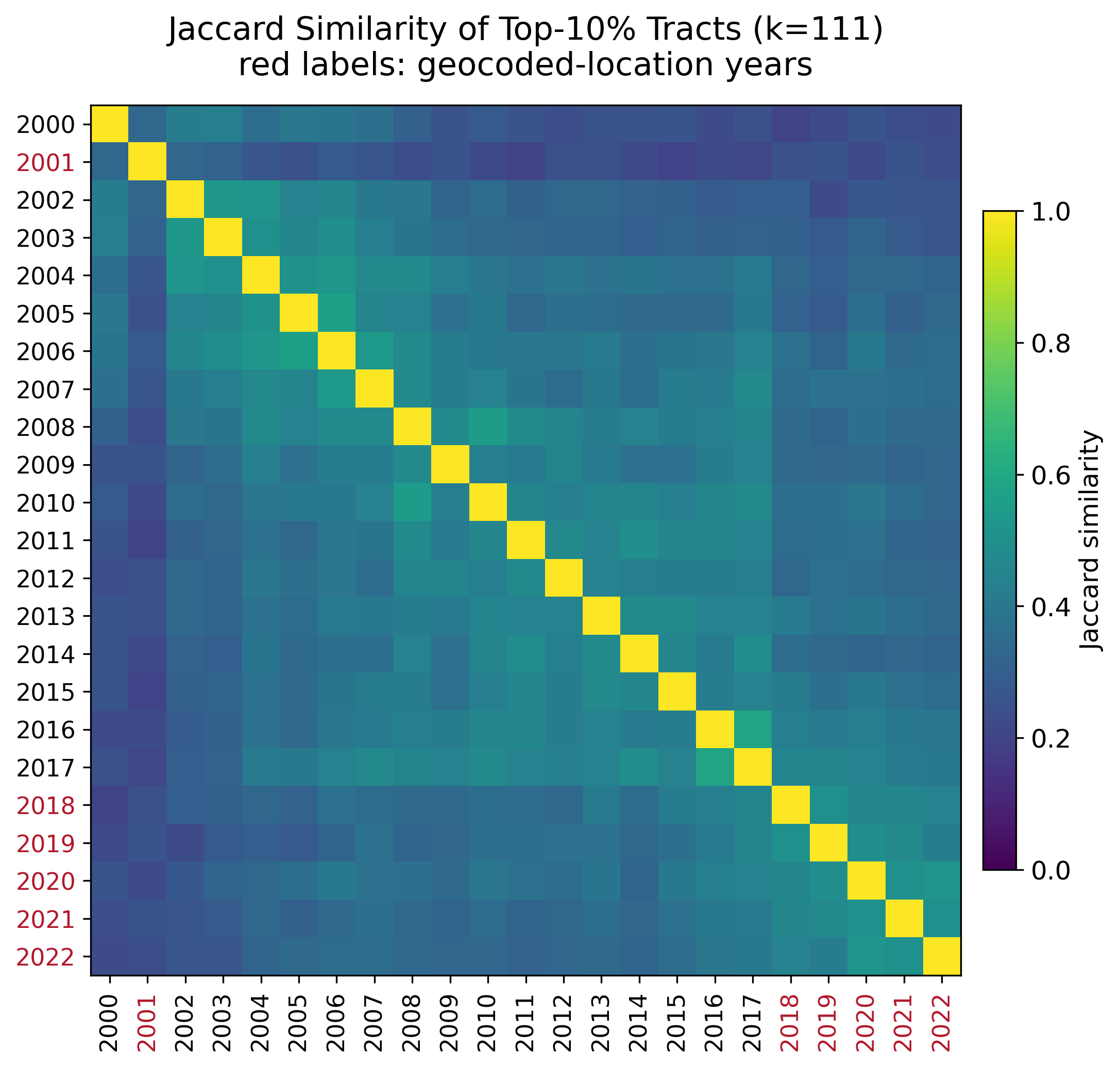}
\caption{Left: years as an annual $G_i^*$ hotspot of the theft rate (FDR $q\le0.05$). Right: Jaccard similarity of the top 10\% of tracts by annual theft count; red labels mark geocoded years.}
\label{fig:persistence}
\end{figure}

\paragraph{Persistence.}
We ran $G_i^*$ separately for each year on the theft rate per 1{,}000 household vehicles (queen contiguity, 9{,}999 permutations, FDR control within each year) for the 1{,}083 tracts with at least 100 household vehicles in every year. Every year has between 10 and 29 hotspots. Six tracts are hotspots in all 23 years and nine in at least 20 (Table~\ref{tab:persistence}); all lie 2--3.5\,km south and southwest of Downtown (Fig.~\ref{fig:persistence}, left). Five of the six are among the 20\% of tracts with the fewest household vehicles, so their high rates partly reflect small denominators. Without FDR control, 20 tracts would qualify in all years. Of the hotspots of one year, 48--100\% remain hotspots in the next.

\begin{table}[t]
\centering
\caption{Number of tracts that are annual rate hotspots in at least $k$ of 23 years ($n=1{,}083$ tracts).}
\label{tab:persistence}
\small
\begin{tabular}{rrr}
\toprule
$k$ & FDR $q\le0.05$ & Uncorrected $p\le0.05$ \\
\midrule
1 & 56 & 187 \\
5 & 20 & 134 \\
10 & 16 & 87 \\
15 & 11 & 53 \\
20 & 9 & 37 \\
23 & 6 & 20 \\
\bottomrule
\end{tabular}
\end{table}

The top 10\% of tracts by annual theft count (111 tracts) changes considerably from year to year (Fig.~\ref{fig:persistence}, right). Consecutive years share a mean Jaccard similarity of 0.49 when both have the same location source, but only 0.37 across a switch of source. The low similarity of 2001 with its neighboring years is therefore partly an artifact of geocoding.

\paragraph{Drift and daily rhythm.}
The center of mass of thefts moved south (Mann--Kendall $\tau=-0.55$, $p=0.009$; $\tau=-0.71$ in years with stored coordinates), with no significant east--west trend. The daily rhythm differs only slightly between persistent hotspots and other tracts (Cram\'er's $V=0.06$), but in a consistent direction: in tracts that are hotspots in at least 15 years, 12.8\% of thefts occur between 07:00 and 09:00 and 56\% are reported on the day they occur, against 5.9\% and 35\% in tracts that are rarely hotspots. This suggests that more thefts in the persistent hotspots happen during the day, for example from parked commuter vehicles, and are noticed soon after.

\section{Vehicle Targeting}
\label{sec:vehicles}

\begin{figure}[t]
\centering
\includegraphics[width=0.85\linewidth]{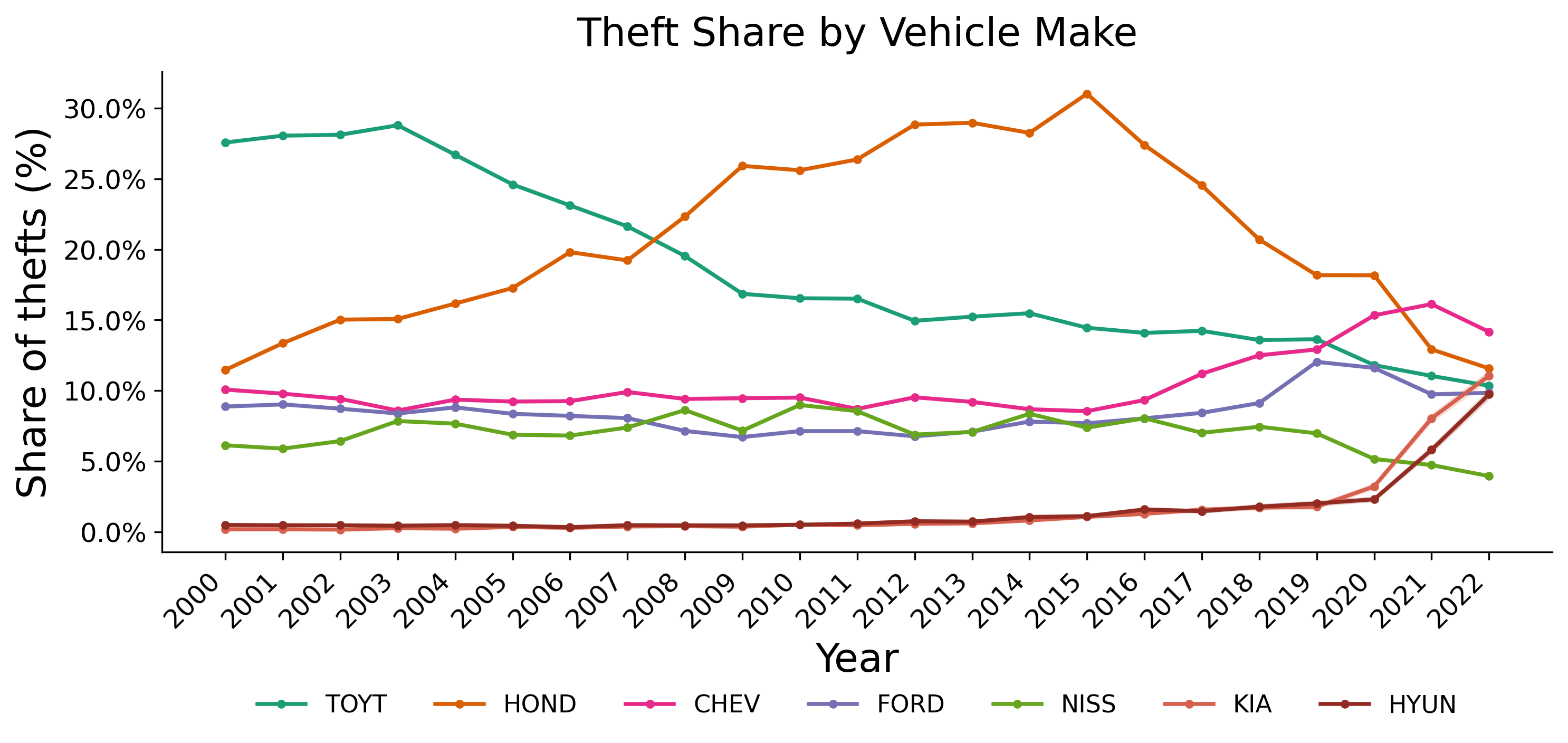}
\caption{Share of thefts by vehicle make, 2000--2022, with 95\% confidence intervals for Kia and Hyundai.}
\label{fig:makes}
\end{figure}

\paragraph{Makes and the Kia and Hyundai wave.}
Toyota (19.7\% of thefts with a known make) and Honda (19.6\%) are the most stolen makes over the whole period, followed by Chevrolet, Ford and Nissan. The largest change is the rise of Kia and Hyundai (Fig.~\ref{fig:makes}): their shares grew from 1.5\% and 1.6\% in 2015--2019 to 9.5\% and 7.7\% in 2021--2022, reaching 11.1\% and 9.7\% in 2022. The rise began in 2020 (Kia 3.2\%) and accelerated in 2021, well before the manufacturers' software campaign of 2023~\cite{nhtsa2023kia}. In 2021 and 2022, 94\% of the stolen Kia and Hyundai vehicles were of model years 2011--2021, in which most of these models lacked a standard immobilizer~\cite{hldi2023hyundaikia}, against 80.5\% in 2019. With LAPD data, Brantingham et al.~\cite{brantingham2026forecasting} also find that the Los Angeles theft wave of these models began before the social media trend became widely known.

\begin{figure}[t]
\centering
\includegraphics[width=0.55\linewidth]{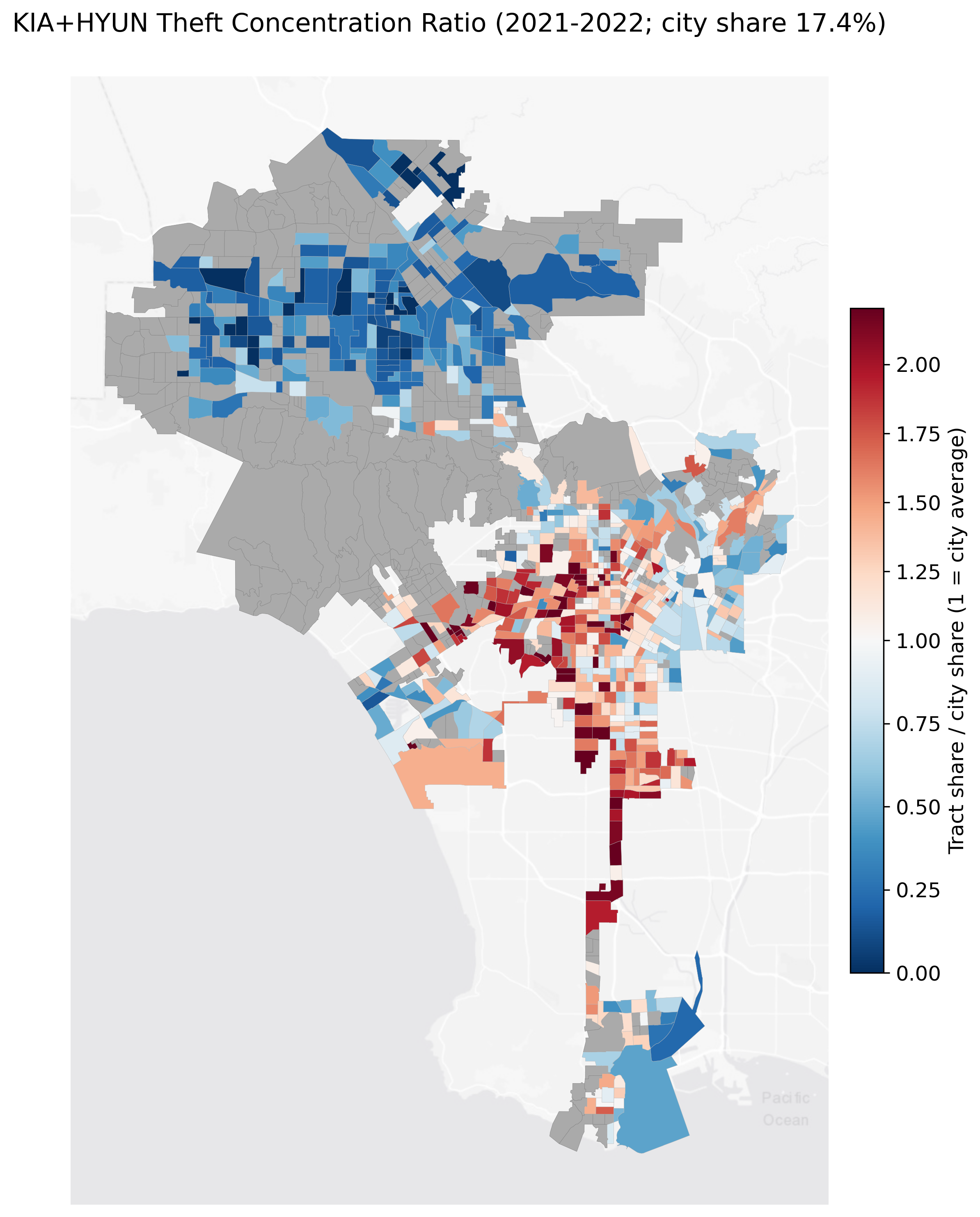}
\caption{Kia and Hyundai share of a tract's thefts relative to the citywide share, 2021--2022; tracts with fewer than 30 thefts are gray.}
\label{fig:kiamap}
\end{figure}

\paragraph{Geography of makes.}
The share of a make among a tract's thefts also reflects the local vehicle fleet, which we do not observe, so these maps describe where makes are stolen rather than targeting. To separate pattern from noise, we compare the number of tracts where a make's share exceeds 1.5 times its citywide share with the number expected from sampling variation alone. For Kia and Hyundai in 2021--2022, 158 tracts exceed this threshold against 45 expected by chance, and the shares are spatially clustered (Moran's $I=0.67$; Fig.~\ref{fig:kiamap}). For Hyundai over the whole period, 220 tracts exceed it against 135 expected, so most of that pattern is noise.

\paragraph{Vehicle age.}
Stolen vehicles were older than the average U.S. light vehicle~\cite{bts_fleetage} in every year. The gap grew from 1.4 years in 2000 to 2.7 years in 2014 and narrowed to 0.6 years in 2022, when vehicles of model year 2010 or later made up 52.6\% of thefts. Median stolen-vehicle age is strongly clustered across tracts (Moran's $I=0.83$). Tracts that are hotspots at $p\le0.05$ have older stolen vehicles than other tracts (median 15 vs.\ 12 years, Mann--Whitney $p<0.001$); for the seven FDR hotspots the difference is not significant.

\section{Discussion}
\label{sec:discussion}

\paragraph{What the results mean.}
Vehicle theft in Los Angeles is concentrated at fine scales, and a small core of high-rate tracts persisted for two decades even while citywide volume halved and recovered, in line with the stability of crime at places reported for Seattle~\cite{weisburd2004trajectories}. Most of these tracts have few residents, so their rates partly reflect small denominators, and the wider set of high-count tracts changes more from year to year. Together with the rising Moran's $I$, this suggests that the conditions producing theft are tied to places more than to periods, which supports long-term, place-based prevention over purely reactive deployment~\cite{braga2019hot}. Short-term space-time clustering exists but is weak beyond a few hundred meters and a week, and half of it consists of repeats at the same location; for prediction, this favors models that combine persistent place effects with short-lived excitation~\cite{mohler2011self}. The Kia and Hyundai wave is consistent with a design weakness in a widely sold fleet changing the composition of thefts within two years.

\paragraph{How preparation shapes results.}
Several patterns that a direct analysis of the raw files would report do not survive preparation. Counting recoveries and traffic reports as thefts inflates volume about sevenfold and shifts the location and timing of events. Parsing the time field naively assigns every 2022 theft to 01:00 and doubles the share of thefts between 00:00 and 02:00, from 5.0\% to 10.0\%. Matching older census data to 2020 tracts by identifier alone leaves 42--60\% of tracts without a value for 2000--2009, and treating missing census values as zero gives 49 populated tracts no vehicles in the 2009 vintage. Changes of location source lower the year-to-year Jaccard similarity of the top tracts from 0.49 to 0.37. We report these effects because similar administrative records are widely used.

\paragraph{Limitations.}
The analysis is descriptive and covers a single city. The accuracy of geocoded locations is validated only for numbered addresses in years with stored coordinates; intersection matches, 18\% of geocoded points in the geocoded years, are not validated. Household vehicles approximate exposure poorly in commercial areas, and reallocating census data by land area can place residents on non-residential land: tract 2774.00 next to LAX receives 1{,}644--4{,}798 residents in older vintages but none in 2022. Tract-level results depend on the spatial units~\cite{openshaw1984maup}, tests across tracts assume independence that does not hold, and the POI data postdate the thefts. Annual 2022 counts are slightly incomplete, recovery figures for recent thefts are censored, and the crime records are not public.

\section{Conclusion}
\label{sec:conclusion}

Analyzing 23 years of LAPD stolen-vehicle reports, we find strong and persistent concentration of theft, weak short-range space-time clustering, no association with monthly temperature, an evening peak, a strong association with lower income at equal exposure, and a rapid shift toward Kia and Hyundai vehicles from 2020 on. The main lesson beyond Los Angeles concerns method: in long administrative records, the definition of events, the handling of time and location fields, and the harmonization of denominators determine many of the patterns one finds. Future work could model the persistent place effects and short-term excitation jointly, for example with self-exciting point processes~\cite{mohler2011self}, and compare cities with different climates and vehicle fleets.

\bibliographystyle{splncs04}
\bibliography{references}

\end{document}